Nataliia Kaliuzhna
https://orcid.org/0000-0003-3154-8194
TIB – Leibniz Information Centre for Science and Technology and University library, Hanover, Germany
Kyiv National University of Culture and Arts, Kyiv, Ukraine
Nataliia.Kaliuzhna@tib.eu

Christian Hauschke,
https://orcid.org/0000-0003-2499-7741
TIB – Leibniz Information Centre for Science and Technology and University library, Hanover, Germany
christian.hauschke@tib.eu

**Open Access in Ukraine: characteristics and evolution from 2012 to 2021**

**Abstract**

This study investigates development of open access (OA) to publications produced by authors affiliated with Ukrainian universities and research organisations in the period 2012-2021. In order to get a comprehensive overview we assembled data from three popular databases: Dimensions, Web of Science (WoS) and Scopus. Our final dataset consisted of 187,135 records. To determine the OA status of each article, this study utilised Unpaywall data which was obtained via API. It was determined that 71.5% of all considered articles during the observed period were openly available at the time of analysis. Our findings show that gold OA was the most prevalent type of OA through a 10 years studied period. We also took a look at how OA varies by research fields, how dominant large commercial publishers are in disseminating national research and the preferences of authors regarding where to self-archive articles versions. We concluded that Ukraine needs to be thoughtful with engagement with large publishers and make sure academics control publishing, not for profit companies, which would monopolise research output distribution, leaving national publishers behind. Beyond that we put a special emphasis on the importance of FAIRness of national scholarly communication infrastructure in monitoring OA uptake.

## 1. Introduction

Open Access as "the practice of providing access to peer-reviewed scientific research articles from all academic disciplines that is cost-free for users and can be reused" (European Commision, 2017) is steadily changing its perception from being a mere alternative to subscription publishing model to a widely acknowledged and integral element of research culture. However, the decades-long battle over how best to achieve comprehensive Open Access implementation continues. It has become evident that successful adoption requires the establishment of sustainable business models and the implementation of robust policies to enforce it (Borrego et al., 2021). Among the latest policies is the mandate by the White House Office of Science and Technology Policy, which requires that publications and their supporting data from federally funded research must be accessible immediately upon publication, with the deadline set for the year 2026 (Horder, 2022). Interestingly, unlike the similar well-known European collective initiative of cOAlition S funders — Plan S — the new US policy does not limit researchers in choosing publication venues: scientists contributing to subscription-based journals could potentially meet the requirement by submitting the peer-reviewed and accepted draft to a publicly accessible repository or another outlet approved by the relevant authority (Brainard & Kaiser, 2022). Of particular relevance is also the European Council "Conclusions on high-quality, transparent, open, trustworthy and equitable scholarly publishing" which calls on the member states to back policies that promote a non-profit and diverse scholarly publishing approach, free for both authors and readers (European Council, 2023).

With the increasing importance of open science and open access on the policy level, it is gaining more importance in research evaluation, too. For instance, starting from 2019 Leiden Ranking[1] introduced open access indicators to their methodology. Another example is the ongoing Open Universal Science Project (OPUS)[2] supported by the European Union, aiming to revamp the research evaluation system that encourages, acknowledges and rewards Open Science practises.

---

[1] CWTS Leiden Ranking https://www.leidenranking.com/
[2] Open and Universal Science (OPUS) Project https://opusproject.eu/

In this context, it is of paramount importance to keep a watchful eye on the expansion of open access literature. In order to effectively accomplish this, the European Commission launched the Open Science Monitor, which aimed to systematically oversee the ongoing transformation to scholarly practices in terms of openness, with particular emphasis on OA across various countries and disciplines. This initiative was soon followed by the establishment of national Open Access Monitors (Barbers, et al., 2022; French Open Science Monitor, 2023). Moreover, governments around the world have strategically aligned their path towards open access by introducing national strategies and action plans tailored to the unique domestic research landscapes and needs of their respective countries. Noteworthy examples include Ireland, Spain, Slovenia, Colombia and Ukraine.

To date, empirical research on open access in Ukraine remains limited. In the scientific literature, there are occasional studies on OA at the level of single universities (Kostyrko & Korolova, 2021) or journals (Yaroshenko & Yaroshenko, 2021). There are also studies on research data sharing practices (Boiko et al., 2021) and black open access analysis (Nazarovets, 2018). Awareness and perceptions on Open Science practices, including open access publishing of Ukrainian researchers, have been partially explored by the Eurodoc study focusing on early career researchers (Berezko et al., 2021). The open access status of papers resulting from funding of the National Research Foundation of Ukraine (Tsiura & Tsybenko, 2022) has been highlighted, too. Finally, the landscape of Ukrainian scientific journals was touched in a multitude of research endeavours (Hladchenko & Moed, 2021; Moed et al., 2021).

Against this background, the aim of this paper is twofold: (1) to fill the gap and explore the open access uptake in Ukraine, its patterns and growth in order to lay a foundation for evidence-based policy development; (2) to enrich the body of literature on open access by illustrating the journey of a country that has cultivated an open access culture without a national OA policy, transformative agreements, or funders mandates.

Our study addresses the following research questions:

    RQ1. How did the share of OA publications produced by researchers affiliated with Ukrainian institutions develop over the period of 2012–2021?

RQ2. What are the characteristics of Ukrainian OA publication fraction?

## 2. Background

The adoption of open access practices in Western and Central European countries differs from the experiences of Eastern European countries. After the collapse of the Soviet Union and the regaining of their independence, the Eastern European countries faced several fundamental challenges related to both socio-economic changes and the reorganisation of their science and research systems. Due to insufficient funding and the inability to purchase resources, the libraries of these countries were pushed to seek alternative ways of commercial publishing and to accumulate scarce resources for the development of local infrastructures that helped to maintain and increase the access to information resources (Donabedian & John Carey, 2011; Johnson, 2014). There were a number of international initiatives that came to help and aimed at fostering development and implementation of electronic systems, improving access to scientific literature, training librarians and promoting the open access movement. For example, the New Independent States of the former Soviet Union (INTAS) Armenia, Azerbaijan, Belarus, Georgia, Kazakhstan, Kyrgyzstan, Moldova, Tajikistan, Turkmenistan, Ukraine and Uzbekistan received access to full-text databases of scientific literature from Springer, Backwell Science, Zentralblatt Mathematik etc. as part of the "Electronic Library" program of the International Association for the promotion of co-operation with scientists (Yaroshenko, 2009). Estonian, Lithuanian and Latvian librarians benefited from the CELIP (Project Central and Eastern European Licensing Information Platform) and increased their knowledge of electronic content licensing (Zumer, 2001). The Electronic Information for Libraries (EIFL) initiative emerged as a prominent advocate, providing substantial support to library consortia in Estonia, Latvia, Lithuania, Poland, Serbia, Slovenia, and Ukraine. Through the allocation of grants, EIFL facilitated the implementation of comprehensive national and institutional open access advocacy campaigns, aimed at engaging and informing research communities about the significance of OA practices (Schmidt & Kuchma, 2012).

In Hungary, for example, the Hungarian Academy of Sciences had already issued its open access mandate in 2013, requiring researchers to make their scientific output freely accessible either by self-archiving or by publishing in gold or hybrid OA journals. Additionally, the Law of Higher Education passed in the same year, requiring that Ph.D. dissertations be made accessible in open access (Karácsony & Görögh, 2017). In Poland, the representative body of academic institutions – the Conference of Rectors of Academic Schools and the Polish Academy of Science fully adapted the European Commission Recommendation on access to and preservation of scientific information (issued on 17 July 2012) in the same year. Within two years, the Polish Ministry of Science and Higher Education issued the guidelines of the OA policy in Poland (Wałek, 2017). The National Science Centre Poland, the largest funding agency for basic research in the country, endorsed Plan S, and also worked on its implementation (Korytkowski & Kulczycki, 2021). Likewise, Slovakia, which joined the European Union in 2004, was firmly set on the path of OA and established the Slovak Center of Scientific and Technical Information to serve as a centre for nationally coordinating OA. Later on, in 2017 the government approved the National Action Plan of the Slovak Republic (Pendse, 2019).

It is important to note that European Union policy has played an important role in driving the implementation of open access in its member states (Karácsony G. & Görögh, 2017). In particular, the PASTEUR4OA project, which aimed at harmonising open access and open data policies within the EU (Picarra, 2015). Other noteworthy examples are Horizon 2020 and Horizon Europe funding schemes with explicit OA mandates. Additionally, the Open Research Europe publishing platform is designed for and limited to researchers funded within the Research and Innovation program. Finally, there are several initiatives, including OpenAIRE and European Open Science Cloud, which have specific sets of requirements and specifications for those who want to participate in them.

The evolution of open access in Ukraine can be traced through a dual perspective: firstly, external efforts and grassroots endeavours of libraries and universities, and secondly, through governmental policies and measures.

## 2.1 Open Access initiatives in Ukraine

As in other countries, Ukrainian academic libraries took an early lead in advancing the availability of freely accessible resources within the country. Supported by donors such as the Kyiv-Mohyla American Foundation, "Vidrodzhennya" Foundations, USAID and the aforementioned EIFL, several initiatives were undertaken to facilitate licensed access to electronic resources, establish institutional repositories, and promote knowledge about open access journals (Yaroshenko, 2011).

Notably, in January 2008, the universities in the border regions of Ukraine, Belarus, and the Russian Federation announced the "Belgorod Declaration on Open Access to Scientific Knowledge and Cultural Heritage", in which the signatories declared their intention "to stimulate the step-by-step development of online open access to scientific knowledge and cultural heritage, the search for legal solutions for the development of existing legal and financial grounds to accelerate the optimal use of open access (The Action Plan of the Principles of the Belgorod Declaration, 2009). Later, in 2012, another declaration, the "Crimean Declaration of Open Access" was announced (Yaroshenko, 2012). Both declarations expressed a preference for green OA, highlighting the significance of institutional repositories in enhancing visibility and recognition of national research. The ideas of openness did not go unnoticed by higher education institutions as well. A coalition of 26 Ukrainian and foreign universities signed the "Olvia Declaration" in 2009 and declared their commitment to foster collaboration and facilitate unrestricted scientific communication through the establishment of institutional repositories and open access journals (Yaroshenko, 2014).

It is also important to note that transforming into electronic publishing Ukraine faced a challenge of recognition of electronic licences. EIFL was the institution that started a dialogue to align the legislative regulation with the use of free public licences in Ukraine. EIFL advocated for revisions to the national copyright law to provide a legal framework for authors to share their results and to enable libraries to meet the needs of their end users (Electronic Information for Libraries, n.d).

## 2.2 Emergence and development of Open Access policies in Ukraine

The first steps towards acknowledging the principles of open access in Ukraine at the state level can be traced back to 2007 with the enactment of the Law of Ukraine "On the Basic Principles of Information Society Development in Ukraine for 2007-2015". This law covered themes like creation of electronic information resources, protection of intellectual property rights, implementation of electronic document management, and enhancing information security. Importantly, it emphasised the need for mandatory storage of research papers in a unified electronic format and ensuring free access to publicly funded research output (Sokolova, 2019; Government of Ukraine, 2007). To achieve this, the "Ukrainian Scientific Periodicals" repository was established. Subsequently, national publishers of academic journals were required to submit full-text articles in HTML or PDF format, along with journal cover and publication metadata to the Vernadsky National Library, which developed and maintained the portal (Government of Ukraine, 2008).

Going one step further, in 2015, a mandate from the Ministry of Education and Science stipulated that dissertations and abstracts must be accessible to the public via institutional websites for three months (Ministry of Education and Science of Ukraine, 2015). This measure was implemented to facilitate public access and promote transparency in scientific research conducted within the country. Subsequently, in January 2022, following the introduction of a new procedure for awarding the Doctor of Philosophy degree, this requirement underwent a revision. A new law required that the electronic version of a dissertation should be archived within a local repository and submitted to the Ukrainian Institute of Scientific and Technical Expertise of Information, when it would then be made available through the National Repository of Academic Texts (NRAT) (Law of Ukraine, 2022). It is important to emphasise that the launch of the National Repository of Academic Texts (NRAT) in 2018 marked a significant milestone in developing national scholarly communication infrastructure. Functioning as a multidisciplinary repository designed for the storage and dissemination of diverse scientific research documents NRAT opened up opportunities to work towards integration with European Open Science Cloud (EOSC) and OpenAIRE (Chmyr, 2019). In 2022, the NRAT granted permission to index its content to anyone and enriched records with metadata needed for publication validation (National Repository of Academic Texts, 2023).

Key issues of OA implementation were further reflected in the updated "Roadmap for integrating Ukraine's research and innovation system into the European Research Area (ERA)". In particular, this document highlighted the need of supporting institutions to cover APCs in international OA journals, establishing agreements with publishers, strengthening national publishing infrastructure and increasing Ukrainian participation in international initiatives and working groups (Go FAIR, DORA, Research Data Alliance etc.) (Ministry of Education and Science of Ukraine, 2021). Continuing this progressive stride towards the embracement of open access practices, Ukraine adopted its National Open Science Action Plan in 2022, declaring its commitment to align national legislation in consonance with European Union standards and norms for data sharing and open access (Government of Ukraine, 2022).

## 3. Methodology

### 3.1 Data collection

In studies examining the share and number of open access publications at a national level, the results could be influenced by the data source chosen and the coverage of publications it indexes. The most comprehensive sources include national bibliographic databases or current research information systems (CRIS), which capture rich metadata on publications that are usually not included in proprietary bibliometric databases, do not have a DOI or are published in a small national journals (Pölönen et al., 2020). The completeness and accuracy of the data provided by a data source as well as the way in which the data is made available and can be reused are of crucial importance, too (Waltman & Larivière, 2020). It would be logical and valuable to use the national Ukrainian scholarly communication infrastructure to study the uptake of OA, but this is not possible for several reasons. First, the National Depository of National Periodics does not provide an API or other means of exporting data, so the metadata is completely locked and not reusable. Second, the two major information sources on institutional publication data — the Ukrainian Research Information System (URIS) and the National Repository of Academic Text — are currently under development and could not be used for data collection either (Kaliuzhna & Auhunas, 2022). Therefore, in order to obtain a detailed profile of OA publications by authors affiliated with Ukrainian institutions, we decided to assemble data from three widely used bibliographic databases: Dimension, Scopus and Web of Science (WoS). The choice of WoS and Scopus is justified by their firm embedding in the Ukrainian research assessment framework. Particularly, since 2013 publications indexed in Scopus and WoS have become a requirement for obtaining a doctoral degree[3]; since 2016 — for granting the academic titles of associate professor and professor. Since 2018, Scopus/WoS indexed articles have been introduced as a criterion for the state certification of Ukrainian higher education institutions. Finally, in 2019, Scopus/WoS publications were included among the criteria for evaluating research projects seeking state funding (Abramo et al., 2023). The Dimension database was selected to complement the afore-mentioned databases because of its broader coverage of smaller national academic journals and higher inclusion of non-English publications in comparison to Scopus and WoS (Visser et al., 2021; Basson et al, 2021).

Data collection took place in November 2022. The search strategy was developed to accommodate each database's features and functionalities. The SciVal web interface was used to extract data from Scopus. An advanced search was performed based on the following query: AFFILCOUNTRY( ukraine ) AND ( LIMIT TO (DOCTYPE, "ar") LIMIT TO (PUBYEAR, 2021) OR LIMIT TO (PUBYEAR, 2020) LIMIT TO (PUBYEAR, 2019) OR LIMIT TO (PUBYEAR, 2018) OR LIMIT TO (PUBYEAR, 2017) OR LIMIT TO (PUBYEAR, 2016) OR LIMIT TO (PUBYEAR, 2015) OR LIMIT TO (PUBYEAR, 2014) OR LIMIT TO (PUBYEAR, 2013) OR LIMIT TO (PUBYEAR, 2012). SciVal's capabilities allow it to export 100,000 publications at a time. Data from WoS (Core Collections) were retrieved via InCites by splitting the query *CU= Ukraine and Article (Document Types)* into two time

---

[3] Prior to January 2022, a prerequisite for obtaining a doctoral degree were publications in Scopus/Wos indexed journals. A new procedure for conferring the Doctor of Philosophy degree accepts publications in national journals as sufficient to fulfil the criteria.

periods from 2012 to 2017 and from 2021 to 2018 to cope with the data export limitation of up to 50,000 records. In the case of Dimensions, the data was extracted via an API that allows advanced research data analysis using a specific Domain Specific Language (DSL). However, there is also a limit for data extraction of up to 50,000 records. In order to gather data for the studied period, the following query was run four times, from 2012 to 2015, from 2016 to 2018, from 2019 to 2020 and for 2021:

*("""search publications*

    *where research_org_country_names = "Ukraine"*

    *and year in [2012 : 2015]*

    *and type = "article"*

*return publications*

*[id+title+doi+year+journal+issn+publisher+times_cited+category_for_2020+category_bra]""").*

Although all searches in the databases were limited to the publication type "articles", it is important to highlight some discrepancies in the classification of the records. In Dimensions, the classification of publication types is broad and consists of 5 content types: article, book, chapter, monograph, proceedings. The type "article" includes "articles from a scientific journal or trade magazine, including news and editorial content"[4]. Scopus has a more detailed approach, dividing publications into 13 categories. "Article" is defined as peer-reviewed original research or opinion. It also includes case reports, technical and research notes and short communications but editorials, letters, reviews and in some cases notes are treated as separate types. The Scopus editorial team is responsible for these classifications (Elsevier, 2023). Finally, WoS offers 40 document types for refining search results, but their availability depends on the type of subscription. The version of WoS being used for this study offers 6 types, including the Early Access type. Article is defined as "report on new and original works, including research papers, brief communications, technical notes, chronologies, full papers, and case reports". Additionally, WoS supports dual classification for some records - articles and proceedings papers whereas InCites only supports one document type per publication[5],[6].

---

[4] What exactly is covered in the "Publications" in Dimensions?
https://dimensions.freshdesk.com/support/solutions/articles/23000018859-what-exactly-is-covered-in-the-publications-in-dimensions-

**3.2 Data processing**

The next step was to merge the retrieved data into a single dataset of 337,849 records. For publications without a DOI, which accounted for 9% (n=30,542) we queried Crossref Metadata Search by article title and were able to retrieve a DOI for 17% (n=5,199) of the publications. Then, duplicate records were removed based on DOI basis using OpenRefine 3.7.3 (OpenRefine, 2023). The dataset comprising 190,911 records was subsequently matched with Unpaywall data using its API to obtain OA status and additional metadata (OA subtype, publisher, presence of a copy in a repository, name of a repository). OA status was obtained for 187,135 publications. The data collection and processing is shown in Figure 1.

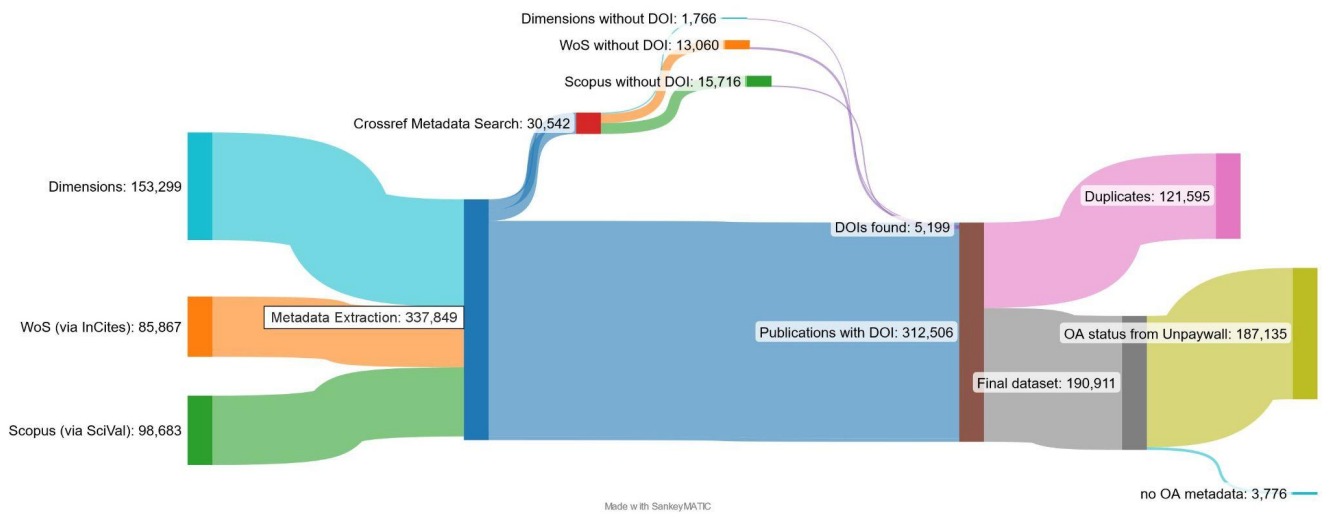

**Figure 1**. Sankey diagram showing the workflow of retrieving, gathering, deduplicating and processing data.

The following definitions for OA subtypes were used in this study to ensure clarity and avoid any ambiguity or misinterpretations:

---

[5] Clarivate, Insites Help. https://incites.help.clarivate.com/Content/Indicators-Handbook/ih-doc-types.htm
[6] Clarivate, Web of Science Help https://webofscience.help.clarivate.com/en-us/Content/document-types.html

- Bronze - articles are free to read on the publisher's website, without a licence that grants any other rights;
- Gold - publication published in a full OA journal;
- Green - articles are published in toll-access journals, but archived in an OA archive, or repository;
- Hybrid - publication freely available under an open licence in a paid-access journal;
- Closed: an OA version of the article has not been found, also referred to as non-OA (Piwowar et al., 2019).

**3.3 Discipline mapping**

To carry out the analysis at the level of disciplines, we mapped those being used in our data sources to the OECD's Fields of Science and Technology classification (FOS) (Stahlschmidt & Stephen, 2020). The revised FOS classification consists of 40 subcategories of 6 major fields, namely Natural Sciences, Engineering and Technology, Medical and Health Sciences, Agricultural Sciences, Social Sciences and Humanities (Revised Fields of Science and Technology, 2007). For Dimension we used the 2-digit code of the [Australian and New Zealand Standard Research Classification (ANZSRC)](#) 2020, for Scopus the 4-digit code of the [All Science Journal Classification Codes (ASJC)](#) and for WoS its native [Research Area scheme](#). Dimensions makes use of machine-learning approaches to classify research objects in its database at the publication object level rather than at the journal level (Porter et al., 2023). Given that Scopus and WoS databases do not provide exclusive research category assignments to individual publications, but rather to journals, publications that received two or more research codes were included in the 7th Multidisciplinary category.

4. Results

**4.1 What is the fraction of open access publications?**

Between 2012 and 2021 Ukraine affiliated authors published 187,135 unique publications. The study revealed a remarkable increase in output, with the number of publications increasing significantly from 7,454 in 2012 to 35,623 in 2021, reaching its highest growth rate of 26,5% in 2015. *Figure 2* displays the number (Fig. 2 A.) and the proportion (Fig. 2 B.) of open access papers published with the overall OA share of 71,5%, ranging from 37,6% (n=2,806) in 2012 to 82% (n=29,197) in 2021. Throughout the study period, the prevalence of paywalled publications showed a steady decline, from 62.4% (n=4,648) in 2012 to 18% (n=6,426) in 2021.

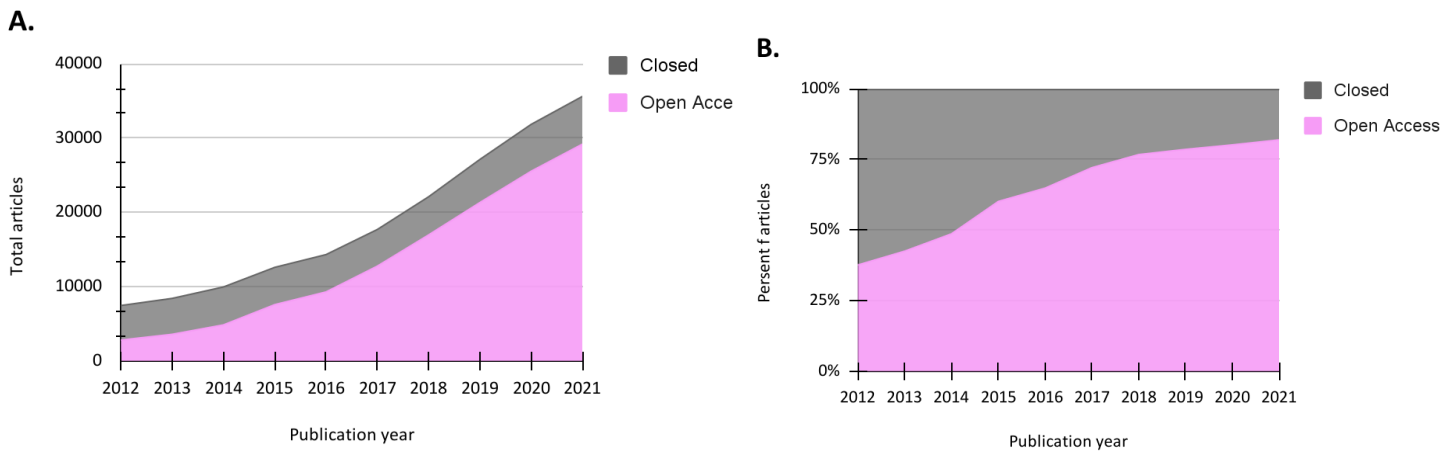

*Figure 2.* Number of articles (A) and proportion of articles (B) with OA papers produced by authors affiliated with Ukrainian institutions from 2012 to 202.

**4.2 What is the prevalence of open access subtypes?**

*Figure 3* illustrates that almost two-thirds of the total number of OA publications 48,1% (n=90,031), are available under the gold OA route. The second most frequent choice was bronze OA, accounting for 12,8% (n=24,046). Hybrid OA and green OA subtypes were the least commonly used with 5,6% (n=10,486) and 5% (n=9,314) of articles published in these categories, respectively. The share of green OA publications dropped from 8,9% (n=667) in 2012 to 3,4% (n=1,208) in 2021, while the prevalence of gold OA increased dramatically from 12% (n=894) in 2012 to 59,8% (n=21,316) in 2021.

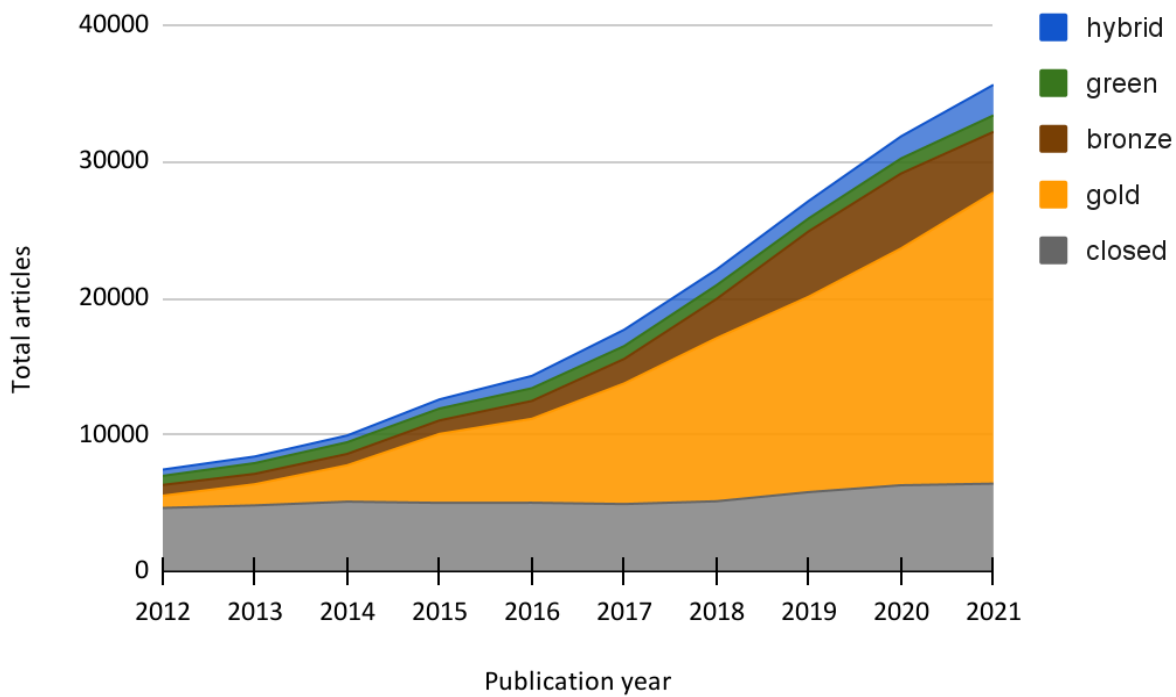

**Figure 3.** Distribution of open access subtypes by publication year

**4.3 How does Open Access vary by research fields?**

Investigation of the OA distribution across different research fields revealed the highest percentage of OA publications in humanities with the rate of 89,7% (n=10,249), closely followed by social sciences at 88,5% (n=34,023). Medical and Health Sciences, Agriculture Sciences and Multidisciplinary categories also exhibited a high level of OA publications, representing 76,4% (n=18,077), 76,1% (n=3,602) and 67,8% (n=10,844) respectively. The lowest proportion of OA papers was observed in natural sciences, with a rate of 55,8% (n=29,883).

*Figure 4.* provides a more granular view on how the different research fields were represented in terms of open access subtypes. Gold OA is dominant in each category, ranging from the highest of 65,4% (n=7,470) in humanities to lowest of 29,8% (n=15,985) in natural sciences. The second largest share is attributed to the bronze subtype, accounting for 20,3% (n=7,809) in social sciences, 15,4% (n=3,646) in medical and health sciences, 14,9% (n=1,703) in humanities, 12,8% (n=2,044) in multidisciplinary, 9,8% (n=5,244) in natural sciences, 8,7% (n=423) in agricultural domain and 5,7% (n=1,474) in engineering and technology. Green OA is the least used way of making papers freely available, showing the highest commitment to self-archiving among researchers

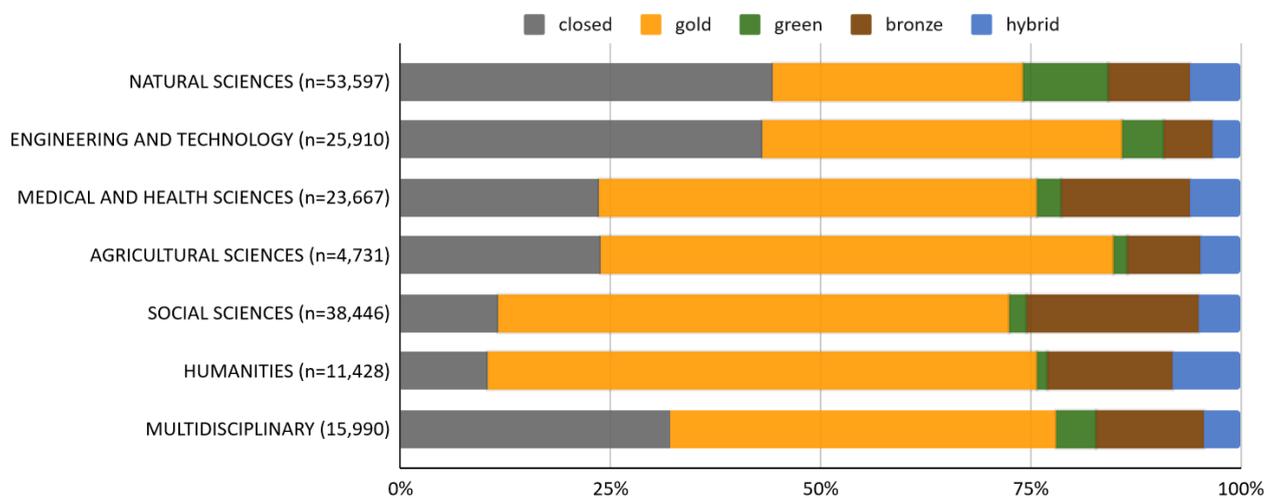

**Figure 4.** Percentage of different open access subtypes per research category.

of natural sciences 10,1% (n=5,399), followed by colleagues of engineering and technology disciplines - 4,9% (n=1,282), multidisciplinary - 4,7% (n=757), while humanities showed the lowest number of archived publications with only 1.2% (n=138).

**4.4 How does open access vary by publisher?**

Publishers were identified for 187,099 publications investigated in our study. *Figure 5* shows that large commercial publishing houses emerged as leaders among publishing houses in distributing Ukrainian research output, with the Springer Science & Business Media published 7,8% (n=14,658) papers of 187,099 articles followed by Elsevier with 6,8% (n=12,744) rate. These publishers' articles were predominantly paywalled, accounting for 70,1% of Springer's and 64,9% of Elsevier's shares respectively. The highest percentage of toll access was observed within Pleiades Publishing (95,3%, n=4,734) being at the 6th position of the most frequently chosen publishers among Ukrainian researchers. 7 out of 10 of the most popular publishers were identified to be national publishers and university presses with a high level of open access publications. For instance, Ukrinform Nauka accounted for 4,9% (n=9,167) and had rate of 42,5% (n=3,899) gold OA, 22,8% (n=2,089) bronze OA, 9,9% (n=907) green OA and 9,1% hybrid OA (n=834). Private Company Technology Center representing 2,5% (n=4,616) had 97,1%(n=6,752) of gold OA and only 1,3% (n=87) closed. University presses of Taras Shevchenko National University of Kyiv, Oles Honchar Dnipropetrovsk National University and Kyiv Polytechnic Institute each represented mostly freely available content. The fractions of paywalled content represented 10,9% for T. Shevchenko National University of Kyiv, 4,8% for both Oles Honchar Dnipro National University and Kyiv Politechnic Institute.

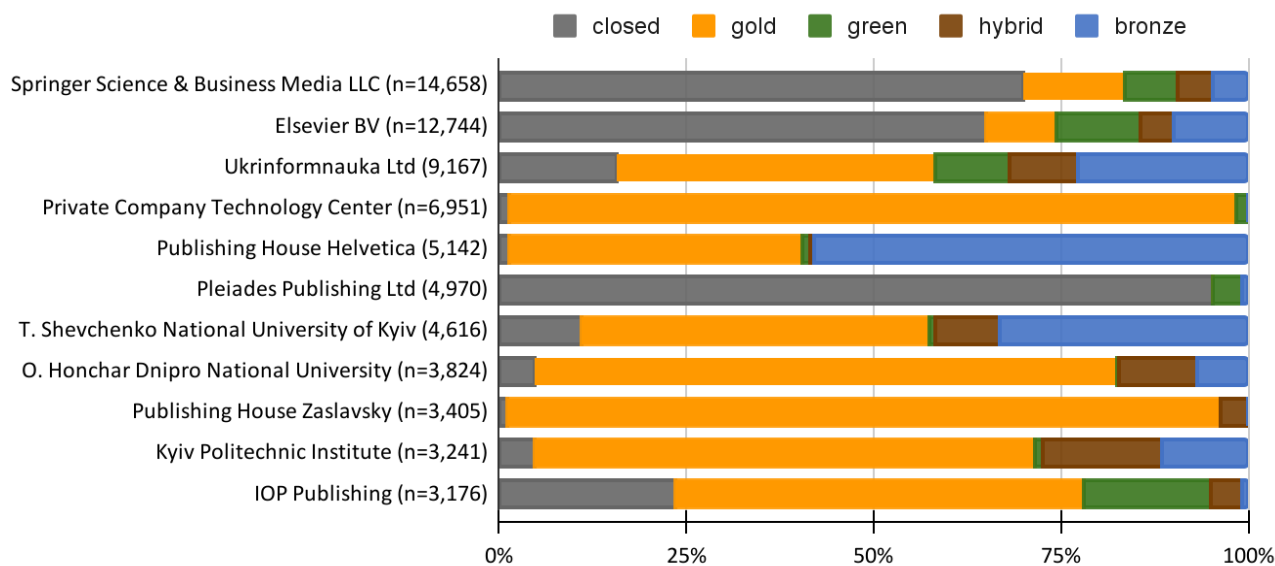

**Figure 5.** Top 10 publishers of distribution research output produced by Ukrainian researchers and open access subtypes provided by them

## 4.5 What are the most popular repositories for self-archiving?

In our dataset, 8,716 publication records possessed metadata on self archiving repositories (*Table 1*). The most preferred choice among Ukrainian researchers for archiving their works was Cornell University's arXiv, which held a significant share of 39.4% (n=3,431). Among the top repositories, 3 out of 10 were national repositories. Notably, "The scientific electronic library of periodicals of the National Academy of Sciences of Ukraine" accounted for 15.2% of the records (n=1,324), followed by Sumy State University Institutional Repository with 4.3% (n=378), and Kharkiv Polytechnic Institute eNTUKhPIIR with 1% (n=87).

Table 1. Top 10 repositories for self-archiving by Ukrainian researchers

| Repository | # of papers | % of papers |
|---|---|---|
| arXiv (Cornell University) | 3,431 | 39,4% |
| The scientific electronic library of periodicals of the National Academy of Sciences of Ukraine | 1,324 | 15,2% |
| Electronic Sumy State University Institutional Repository | 378 | 4,3% |
| Europe PMC | 368 | 4,2% |
| HAL -Hyper Articles en Ligne | 200 | 2,3% |
| RePEc: Research Papers in Economics | 90 | 1% |
| eNTUKhPIIR (Kharkiv Polytechnic Institute ) | 87 | 1% |
| Repository of the Academy Library (Library of the Hungarian Academy of Sciences) | 87 | 1% |
| Red Federada de Repositorios Institucionales de Publicaciones CientÃficas - LA Referencia | 73 | 0,8% |
| MPG.PuRe (Max Planck | 63 | 0,7% |

| Society) | | |
| Other | 2615 | 30% |

## 5. Discussion and conclusion

This study represents the first comprehensive empirical investigation of the adoption of open access publications in Ukraine. Based on our dataset, the share of OA publications at the national level from 2012 to 2021 was 71,5%, which is a high rate considering the absence of an established national OA policy, funders mandates and transformative agreements within the country. Notably, this percentage is significantly higher than the global average of 31% reported by Piwowar et al (2019), as well as their projected estimate of 44% by 2025. In fact, Ukraine is steadily approaches the leading nations in Europe regarding their share of OA publications, very close to Great Britain with the median share of OA publication of 74% in 2020 (Robinson-Garcia et al., 2020) and the Netherlands with 89% in 2022 (Puylaert & Kooistra, 2023). It is worth mentioning that the COKI Open Access Dashboard, which utilises data from OpenAlex and Crossref reports different levels of OA for all countries, for example 53% for Ukraine and 54% for the Netherlands (Diprose et al., 2023). This can be explained by usage of different data sources and OA classification.

Our study further substantiates the phenomenon observed by Maddi et al. (2021), which highlights the high adoption of open access in some Eastern European countries. This success can be attributed primarily to the active involvement of institutions in advocating for open access principles. Regarding the distribution of the different subtypes of OA, our analysis revealed that gold OA was the dominant category, constituting 48,1% of the total articles. Furthermore, the growth of this particular subtype was observed consistently growing throughout the entire study period, ranging from 12% of all OA papers in 2012 to 59,8% in 2021. This tendency can be explained by the fact that Ukrainian researchers tend to publish their papers predominantly in national OA journals. Nazarovets (2020) found that one third of papers produced by researchers affiliated with Ukrainian research institutions and universities indexed in Scopus (2015-2019) were published in Ukrainian outlets. Similarly, based on a study of 48 selected Scopus-indexed Ukrainian journals Hladchenko & Moed (2021) revealed that Ukrainian authors constituted 68.2% of publications in these journals. The rationale for selecting national journals may vary, it would be valuable to perform empirical models to identify them. However, the existing extensive network of OA journals most likely plays a contributory role in this regard (Khanna, 2022).

It is generally assumed that OA in humanities is not as prevalent as it is in different fields. Archambault et al. (2014) report a share of 31.3 % for general arts, humanities & social sciences from 2011 - 2013, which belongs to the lowest shares in this study. On a national level this has been researched for different countries. A similar share of 28 % to 32 % for the years from 2011 to 2014 was found in Spain (Torres Salinas et al. 2016) and 29.8% in Finland in the arts and humanities for 2016-2017 (Pölönen et al. 2020). A newer study of OA shares on a global level by Simard et al. (2021) calculated 20 % for social sciences and humanities. Although these studies have different methodologies and target areas, they reach relatively similar results. All the more remarkable is the very high percentage of Open Access in the humanities in Ukraine, where it turned out to be the most open domain with a fraction of 89,7 %, closely followed by social sciences with 88,5 %.

There are some commonly cited reasons for a low OA share in these fields, which can be divided into structural or socio-economic on the one hand and cultural barriers on the other hand. Firstly, it is stated that it is more difficult to allocate fundings to covering APCs, and there are limited OA journal options in particular research areas, furthermore lack of time finding out more about OA and frustration in adding journal articles to the institutional repositories (Quigley, 2021). The cultural issues include assertions that OA could encourage plagiarism-like practices, and concerns regarding quality control of oa content (Mandler, 2013). In Ukraine it's different because Ukrainian researchers from these fields predominantly publish their research output in Ukrainian language in national journals. Most Ukrainian journals are Diamond OA journals or charge an affordable APC (Novikov, 2020). If researchers want to publish their articles in their own language, it will most likely be OA by default, even though it leads to less international recognition (Nazarovets & Mryglod, 2023).

A notable difficulty was the selection of data sources for such an analysis. Being aware of limited and biassed coverage of Ukrainian research by proprietary WoS and Scopus we sought the possibility to reuse national publication data provided by domestic bibliographic sources. The current state of the Ukrainian scholarly communication infrastructure is characterised by partial non-compliance with the FAIR principles for research information on the one hand, and the ongoing development of its new blocks on the other hand, what hindered the task of leveraging scholarly metadata on Ukrainian research output, its assessment and monitoring (Hauschke et al., 2021). The development of a national research information system with a module for OA monitoring offers potential to significantly improve the situation. The data sources used for this investigation due to the lack of a national CRIS have their own shortcomings. Specifically, different definitions of research output categories. Another big challenge in the data collection and processing was the inconsistent and incomplete usage of country affiliations of authors in the chosen data sources. Mryglod & Nazarovets (2023) investigated this phenomenon for Scopus and WoS. Our experiences indicate that Dimension metadata have the same or a similar issue, further research is needed.

The conclusions drawn from our study indicate the important role of Ukrainian scholarly-led publishing infrastructures in fostering dissemination of national research in a freely available way. Future research needed to investigate institutional spending on OA publication fees in Ukraine. This would bring more clarity on whether it is reasonable for Ukraine to engage into negotiation with large commercial publishers for "read and publish" agreements. Such agreements have been proven to be effective mechanisms for increasing the share of open access (OA) research, however, they are not sustainable and can severely impact market competition, potentially leaving national publishers behind (Haucap et al., 2021). In the case of Ukraine, it would literally mean voluntarily transferring control of national research output to publishing corporations, while a number of countries are tirelessly working on strategies for going beyond transformative agreements and regain control of research output within academia.

Yet, the ongoing war waged by the Russian Federation against Ukraine cannot be overlooked in this context, too, as it significantly underscores the detrimental effects on the scientific domain. Recent research has already highlighted a decrease in the number of publications and a decline in research collaboration networks among Ukrainian authors (Damaševičius & Zailskaitė-Jakštė, 2023). While a number of major international publishers have waived APCs for Ukrainian authors, it remains challenging for researchers to sustain their scientific work due to various factors related to the ongoing conflict. These include the loss of funding from the National Research Foundation of Ukraine (NRFU), the mental strain affecting researchers ability to focus on research, and instances where researchers have had to make the difficult choice of leaving academia to volunteer or join the armed forces of Ukraine (Fiialka, 2022). In a separate investigation, Ukrainian editorial personnel of scientific publishers reported notable impediments in organisation of their editorial workflows. Specifically, 50% of the respondents reported encountering difficulties in this regard, while 3.2% of participants reported temporary suspension of publishing or cessation of publications (Zhenchenko, 2023).

In this context, it is worth noting initiatives launched to keep Ukrainian scholarly publishing afloat, particularly Supporting Ukrainian Editorial Stuff (SUES), which provided both mentoring and financial support, Supporting Ukrainian Publishing Resilience and Recovery (SUPRR) initiative to support Ukrainian publishing now and after the war, Electronic Preservation Project for Ukrainian OA journals (EPP UA), aiming at creating digital archive of Ukrainian OA journals in natural sciences (Stoddard, 2022; Quiñones, 2023; Augunas et al., 2023)

**Limitations**

This study has a number of limitations. Firstly, it focuses exclusively on the publication type "articles," not taking into account other research output types. This exclusion may have introduced potential influences on the results that were not accounted for in this analysis. However, due to the fact that WoS has a dual research types classification for some records, conference proceedings can be included into our data set. Thirdly, the scope of this study does not encompass an examination of the possible impact and extent of OA initiatives mentioned in the paper. Overall, these limitations should be considered when interpreting the results and applying ongoing efforts related to the implementation of the Open Science plan in Ukraine.

**Acknowledgments**

We thank Marco Tullney for shaping this study through valuable discussions. We thank Ludo Waltman for advice on discipline mapping. We are grateful to Iryna Kuchma for providing the original text of the Belgorod Declaration on Open Access. First author is also grateful to Franziska Altemeier and Sabina Auhunas for their moral support.

**Supplementary data:**

Nataliia Kaliuzhna. (2023). Open Access in Ukraine: characteristics and evolution from 2012 to 2021: supplementary data [Data set]. Zenodo. https://doi.org/10.5281/zenodo.8313257

6. **Notes**

[1] CWTS Leiden Ranking https://www.leidenranking.com/

[2] Open and Universal Science (OPUS) Project https://opusproject.eu/

[3] Prior to January 2022, a prerequisite for obtaining a doctoral degree were publications in Scopus/Wos indexed journals. A new procedure for conferring the Doctor of Philosophy degree accepts publications in national journals as sufficient to fulfil the criteria.

[4] What exactly is covered in the "Publications" in Dimensions?

https://dimensions.freshdesk.com/support/solutions/articles/23000018859-what-exactly-is-covered-in-the-publications-in-dimensions-

[5] Clarivate, Insites Help.

https://incites.help.clarivate.com/Content/Indicators-Handbook/ih-doc-types.htm

[6] Clarivate, Web of Science Help

https://webofscience.help.clarivate.com/en-us/Content/document-types.html